\documentclass[twocolumn, tighten]{aastex631}
\usepackage{lineno}

\usepackage{nicefrac}
\usepackage{microtype}
\usepackage{booktabs}
\usepackage{rotating}
\usepackage{xcolor}

\newcommand{\lya}{Ly$\alpha$}

\newcommand{\kms}{${\rm km\,s}^{-1}$}

\newcommand{\FBQS}{TXS 0206$-$048}

\received{?}
\revised{?}
\accepted{?}
\submitjournal{ApJL}

\shorttitle{Tracing filamentary accretion over $>$100 kpc into the ISM of a quasar host at $z\approx1$}
\shortauthors{Johnson et al.}

\begin{document}

\title{Directly tracing cool filamentary accretion over $>100$ kpc into the interstellar medium of a quasar host  at $z\approx1$}

\author[0000-0001-9487-8583]{Sean D. Johnson}
\affil{Department of Astronomy, University of Michigan, 1085 S. University, Ann Arbor, MI 48109, USA}
\correspondingauthor{Sean D. Johnson}
\email{seanjoh@umich.edu}

\author[0000-0002-0668-5560]{Joop Schaye}
\affil{Leiden Observatory, Leiden University, PO Box 9513, NL-2300 RA Leiden, the Netherlands}

\author[0000-0002-6313-6808]{Gregory L. Walth}
\affil{IPAC, California Institute of Technology, Mail Code 314-6, 1200 E. California Blvd., Pasadena, CA 91125}

\author[0000-0002-0311-2812]{Jennifer I-Hsiu Li}
\affil{Department of Astronomy, University of Michigan, 1085 S. University, Ann Arbor, MI 48109, USA}

\author[0000-0002-8459-5413]{Gwen C. Rudie}
\affil{The Observatories of the Carnegie Institution for Science, 813 Santa Barbara Street, Pasadena, CA 91101, USA}

\author[0000-0001-8813-4182]{Hsiao-Wen Chen}
\affil{Department of Astronomy \& Astrophysics, The University of Chicago, Chicago, IL 60637, USA}

\author[0000-0002-8739-3163]{Mandy C. Chen}
\affil{Department of Astronomy \& Astrophysics, The University of Chicago, Chicago, IL 60637, USA}

\author{Beno\^{i}t Epinat}
\affil{Aix Marseille Univ, CNRS, CNES, LAM, Marseille, France}
\affil{Canada-France-Hawaii Telescope, CNRS, 96743 Kamuela, Hawaii, USA}

\author[0000-0003-2754-9258]{Massimo Gaspari}
\affil{Department of Astrophysical Sciences, Princeton University, 4 Ivy Lane, Princeton, NJ 08544-1001, USA}

\author[0000-0001-5804-1428]{Sebastiano Cantalupo}
\affil{Department of Physics, University of Milan Bicocca, Piazza della Scienza 3, I-20126 Milano, Italy}

\author[0000-0002-0417-1494]{Wolfram Kollatschny}
\affiliation{Institut f\"{u}r Astrophysik, Universit\"{a}t G\"{o}ttingen,
Friedrich-Hund Platz 1, 37077 G\"{o}ttingen, Germany}

\author[0000-0002-2662-9363]{Zhuoqi (Will) Liu}
\affil{Department of Astronomy, University of Michigan, 1085 S. University, Ann Arbor, MI 48109, USA}

\author[0000-0003-3938-8762]{Sowgat Muzahid}
\affil{Inter-University Centre for Astronomy and Astrophysics (IUCAA), Post Bag 4, Ganeshkhind, Pune 411 007, India}





\begin{abstract}
We report the discovery of giant (50$-$100 kpc) [O\,II] emitting nebulae with the Multi-Unit Spectroscopic Explorer (MUSE) in the field of \FBQS, a luminous quasar at $z=1.13$. Down-the-barrel UV spectra of the quasar show absorption at velocities coincident with those of the extended nebulae, enabling new insights into inflows and outflows around the quasar host. One nebula exhibits a filamentary morphology extending over 120 kpc from the halo toward the quasar and intersecting with another nebula surrounding the quasar host with a radius of 50 kpc. \textcolor{black}{This is the longest cool filament observed to-date and arises at higher redshift and in a less massive system than those in cool-core clusters}. The filamentary nebula has line-of-sight velocities $>300$ \kms\ from nearby galaxies but matches that of the nebula surrounding the quasar host where they intersect, consistent with accretion of cool inter- or circum-galactic medium or cooling hot halo gas. The kinematics of the nebulae surrounding the quasar host are unusual and complex, with redshifted and blueshifted spiral-like structures. The emission velocities at 5$-$10 kpc from the quasar match those of inflowing absorbing gas observed in UV spectra of the quasar. Together, the extended nebulae and associated redshifted absorption represent a compelling case of cool, filamentary gas accretion from halo scales into the extended interstellar medium and toward the nucleus of a massive quasar host. The inflow rate implied by the combined emission and absorption constraints is well below levels required to sustain the quasar's radiative luminosity,  suggesting anisotropic or variable accretion.


\end{abstract}



\section{Introduction} \label{section:intro}

Observed scaling relations between the interstellar medium (ISM) and star formation in massive star-forming galaxies imply ISM depletion times of a few Gyr at low-$z$ (for a review, see \citealt{Kennicutt:2012}; cf. \citealt{Leitner:2011}) and under 1 Gyr at $z>1$ \citep[e.g.][]{Tacconi:2013}. These ISM depletion timescales are smaller than the age of the Universe, indicating that galaxies must accrete gas from external reservoirs to enable future star formation and black hole growth.  This fresh material can be supplied via mergers \citep[e.g.][]{Moreno:2021}, accretion of cool ($T\lesssim 10^5$ K) inter-/circum-galactic medium \citep[IGM/CGM; e.g.,][]{Dekel:2009}, or cooling of hot  ($T\gtrsim 10^6$ K) gaseous halos \citep[e.g.,][]{Correa:2018}. Despite the importance of gas accretion in galaxy evolution, direct and unambiguous observations of accretion are rare \citep[for a review, see][]{Putman:2017}.

The Milky Way represents a unique case where we can study gas accretion onto a galaxy over many phases and angles. The kinematics of the Milky Way's extraplanar gas exhibit clear signs of accretion in both neutral and ionized phases \citep[for a review, see][]{Putman:2012}, with total inferred mass inflow rates consistent with the Galaxy's star formation rate \citep[e.g.][]{Lehner:2011}.

Beyond the Milky Way, most studies of inflows rely on absorption features in galaxy spectra that are redshifted relative to the galaxy systemic velocity \citep[for a review, see][]{Rubin:2017}. Despite the difficulty in detecting redshifted gas in ``down-the-barrel'' spectra,
surveys at $z\lesssim 1.5$ have revealed likely inflows for ${\approx}80$ galaxies \citep[e.g.][]{Sato:2009, Krug:2010, Coil:2011, Rubin:2012, Martin:2012}. However, the locations of the inflowing gas relative to the galaxy are not directly constrained, leaving their origins and fate unknown.

In contrast, the background absorption spectroscopy often used to characterize the CGM \citep[for reviews, see][]{Chen:2017, Tumlinson:2017} informs the spatial distribution of the CGM and IGM around galaxies but carries little direct information on the radial direction of gas flows for individual systems. In rare cases, absorption spectroscopy can help differentiate inflows from outflows through modeling of velocity shear in multi-sightline data \citep[e.g., ][]{Chen:2014, Lopez:2018}, detailed metallicity gradients \citep[e.g.][]{Fu:2021}, or correspondence with galactic rotation \citep[e.g.][]{Ho:2017, Zabl:2019}, but all of these cases require model assumptions.  Observations that carry direct information on both the morphology and radial direction of gas flows in the same systems have the potential to significantly improve our understanding of accretion onto galaxies.

Wide-field integral field spectrographs (IFS) such as MUSE \citep[][]{Bacon:2010} and KCWI \citep[][]{Martin:2010} enable unprecedented morphological and kinematic maps of giant IGM/CGM nebulae through H\,I Ly$\alpha$ emission at $z\gtrsim2$ \citep[e.g.][]{Cantalupo:2014, Borisova:2016, Cai:2019, OSullivan:2020, Chen:2021, Fossati:2021} and rest-frame NUV$-$optical emission features at lower redshift \citep[][]{Epinat:2018, Johnson:2018, Boselli:2019, Rupke:2019, Chen:2019a, Burchett:2021, Helton:2021, Zabl:2021, Leclercq:2022}.  At $z\lesssim1$, IFS provide highly complete galaxy redshift surveys and access to non-resonant emission lines to directly trace ionized gas morphology and kinematics. To date, IFS data enabled the discovery of halo-scale nebulae arising from large-scale outflows, cool intragroup medium, and stripping of ISM during galaxy interactions including two nebulae with additional insights from nearby intervening absorption spectroscopy \citep[][]{Chen:2019a, Zabl:2021}. \textcolor{black}{Moreover, wide-field IFS enable measurements of the velocity structure function which provide unique insights into turbulence in diffuse gas \citep[e.g.,][]{Li:2020, Chen:2022}.} However, observations of gas accretion with wide-field IFS remain ambiguous with the exception of a giant \lya\ nebulae at $z=3.3$ with evidence of infall coming from morphology \citep[][]{Rauch:2011, Rauch:2016} and the \lya\ emission profile \citep[][]{Vanzella:2017}.

Here, we report the emission detection of an accreting gas filament extending over $>120$ proper kpc (pkpc) from the halo into the ISM around \FBQS, a luminous quasar at $z\approx1$. An archival down-the-barrel UV spectrum of the quasar breaks the inflow/outflow degeneracy that limits intervening absorption studies and reveals inflowing absorption at velocities similar to the nearby nebular emission. The Letter proceeds as follows. In Section \ref{section:observations}, we describe the MUSE, {\it Hubble Space Telescope } ({\it HST}), and Magellan observations and data reduction. In Section \ref{section:quasar}, we characterize the properties of \FBQS \ and its host group. In Section \ref{section:nebulae}, we describe the giant nebulae in the quasar environment and the coincidence with inflowing gas seen in down-the-barrel absorption. In Section \ref{section:conclusions}, we summarize our findings and discuss their implications.  All magnitudes are given in the AB system. Throughout, we adopt a $\Lambda$ cosmology
with $\Omega_{\rm m}\,{=}\,0.3$, $\Omega_\Lambda\,{=}\,0.7$,
and $H_0\,{=}\,70\,{\rm km\,s^{-1}\,Mpc^{-1}}$.

\section{Observations, data reduction, and galaxy measurements}
\label{section:observations}

Investigators studying the IGM and CGM \citep[e.g.,][]{Tejos:2014, Finn:2016} obtained high quality UV absorption spectra of \FBQS\ with the Cosmic Origins Spectrograph \citep[COS;][]{Green:2012} due to its UV brightness, long redshift pathlength, and availability of archival data from the Gemini Deep Deep Survey \citep[GDDS;][]{Abraham:2004}. We obtained the COS G130M and G160M spectra of \FBQS\ (PI: Morris; PID: 12264) from MAST and reduced them following procedures described in \cite{Johnson:2013} and \cite{Chen:2018} to improve the wavelength calibration.

The field near \FBQS\ was observed with the Advanced Camera for Surveys (ACS) aboard {\it HST} with the F814W filter by the GDDS survey for a total of 32 ksec (PI: Abraham; PID: 9760) and for 7.2 ksec by a Sagittarius Stream program (PI: van der Marel; PID 12564), but the quasar falls just $4$ arcsec from the edge of the field-of-view (FoV). To identify and measure the morphologies of faint galaxies near the quasar sightline, we obtained an additional 2.1 ksec of exposure with ACS$+$F814W (PI: L. Straka; PID: 14660). \textcolor{black}{We reduced and coadded the ACS imaging for the field using DrizzlePac \citep[][]{Hoffmann:2021} including Tweakreg for alignment and Astrodrizzle to combine them.} The effective wavelength of ACS$+$F814W corresponds to a rest-frame wavelength $\approx 3700$ \AA\ at $z=1.13$.

To provide deep galaxy redshift survey data, we acquired a total of 8 hours of exposure under $0.7$ arcsec seeing conditions on the field of \FBQS\ with MUSE as part of the MUSE-QuBES survey (PI: J. Schaye, PID: 094.A-0131, 094.A-0131). We reduced the data using the GTO pipeline \citep[][]{Weilbacher:2014} and sky subtraction tools \citep[][]{Soto:2016} as described in \cite{Johnson:2018}. To ensure robustness, we also reduced the MUSE data using CubEx \citep[][]{Borisova:2016, Cantalupo:2019} and the ESO pipeline \citep[][]{Weilbacher:2020} and found consistent conclusions. We scaled the variance array reported by the GTO pipeline by a factor of $1.4$ to better reflect empirical variance \cite[see][]{Herenz:2017}.

To detect faint emission near the quasar, we performed quasar light subtraction as described in \cite{Helton:2021}.  We then identified galaxies in the MUSE FoV by running source extractor \citep[][]{Bertin:1996} on both the ACS image of the field and a median image formed from the MUSE datacube. Finally, we extracted spectra using MPDAF \citep[][]{Piqueras:2017}.

While the MUSE FoV is wide for an IFS, its FoV radius corresponds to a projected distance of $d=250-350$ pkpc from the quasar at $z=1.13$. We supplemented the MUSE data with a wider-field galaxy redshift survey using the LDSS3 spectrograph on Magellan following procedures outlined in \cite{Johnson:2019}. The LDSS3 FoV extends to $d\approx 1.5$ proper Mpc at the $z=1.13$.

For both the MUSE and LDSS3 spectra, we measured redshifts by fitting the observed spectra with linear combinations of the first four galaxy eigenspectra from \cite{Bolton:2012} as described in \cite{Johnson:2018} and \cite{Helton:2021}. \textcolor{black}{To prevent spatially coincident extended nebular emission from biasing galaxy redshifts, we measured galaxy redshifts based purely on stellar absorption \textcolor{black}{when possible} by masking strong emission lines. For galaxies without sufficient continuum signal-to-noise, we measured redshifts with the [O\,II] doublet.} Finally, for galaxies with secure spectroscopic redshifts, we measured absolute magnitudes in the rest-frame $B$-band and 4000 \AA\ break strength \citep[$D_n(4000)$;][]{Balogh:1999}.

\section{Quasar Properties \& environment}
\label{section:quasar}

\FBQS\ is a luminous, core-dominated radio-loud quasar \citep[][]{Becker:2001} at a redshift of $z=1.1317\pm0.0002$ based on a measurement of the [O\,II] emission-line centroid in its MUSE spectrum \textcolor{black}{and adopting the rest-frame effective [O\,II] doublet centroid and systemic uncertainty from \cite{Hewett:2010}}. To estimate the quasar's luminosity and black hole mass, we fit the MUSE quasar spectrum near the Mg\,II emission line with a power-law continuum, Fe\,II template, and 3 Gaussian emission-line components using PyQSOFit \citep[][]{Guo:2019}. The measured monochromatic continuum luminosity at rest-frame 3000 \AA\ implies 
a bolometric luminosity of $\log L_{\rm bol}/{\rm erg\,s^{-1}}\approx 47.2$ using bolometric corrections from \cite{Richards:2006}. The 
Mg\,II line width and luminosity result in an inferred black hole mass of $\log M_{\rm BH}/M_\odot \approx 9.6$ using the single epoch virial theorem estimator from \cite{Shen:2011}.

\begin{table*}
\caption{Summary of galaxies in the field of \FBQS\ at $z\,{\approx}\,z_{\rm QSO}$.}
\label{table:galaxies}
\hspace{-3cm}
\begin{tabular}{rcccccrrrccc}
\hline
\multicolumn{1}{c}{ID} & R.A. & Decl. & $m_{\rm F814W}$ & redshift & redshift & \multicolumn{1}{c}{$\Delta \theta$} & \multicolumn{1}{c}{$d$} & \multicolumn{1}{c}{$\Delta v$} & $D_n(4000)$ & $M_B$ & $\log L_B/L_*$\tablenotemark{a}  \\
     & (J2000) & (J2000) & (AB) & & type & \multicolumn{1}{c}{($''$)} & \multicolumn{1}{c}{(pkpc)} & ($\rm km\,s^{-1}$) &  &  (AB) &  \\
\hline
quasar & 02:09:30.77 & $-$04:38:26.1  &   $-$               &  $1.1317 \pm 0.0002$ & [O\,II]  &  $ 0.0$ & $  0$  &  $ 0$    &    --             &     --      &   --          \\
 G1    & 02:09:30.42 & $-$04:38:29.8  &   $23.82 \pm 0.02$  &  $1.1275 \pm 0.0002$ & stellar  &  $ 6.3$ & $ 51$  &  $ -590$ &   $1.31 \pm 0.05$ &    $-20.4$  &  $-0.5$    \\
 G2    & 02:09:30.92 & $-$04:38:18.6  &   $25.26 \pm 0.08$  &  $1.1313 \pm 0.0004$ & stellar  &  $ 7.9$ & $ 65$  &  $ -60$  &   $1.13 \pm 0.08$ &    $-18.9$  &  $-1.0$    \\
 G3    & 02:09:30.59 & $-$04:38:18.4  &   $23.34 \pm 0.03$  &  $1.1296 \pm 0.0002$ & stellar  &  $ 8.2$ & $ 68$  &  $ -300$ &   $1.06 \pm 0.03$ &    $-20.6$  &  $-0.4$    \\
 G4    & 02:09:31.35 & $-$04:38:27.7  &   $25.77 \pm 0.13$  &  $1.1397 \pm 0.0002$ & [O\,II]  &  $ 8.8$ & $ 72$  &  $+1130$ &   $1.17 \pm 0.25$ &    $-18.4$  &  $-1.2$    \\
 G5    & 02:09:30.14 & $-$04:38:27.1  &   $25.43 \pm 0.09$  &  $1.1365 \pm 0.0002$ & [O\,II]  &  $ 9.3$ & $ 77$  &  $ +680$ &   $0.76 \pm 0.21$ &    $-18.5$  &  $-1.2$    \\
 G6    & 02:09:30.90 & $-$04:38:15.5  &   $24.26 \pm 0.10$  &  $1.1295 \pm 0.0002$ & stellar  &  $10.8$ & $ 89$  &  $ -310$ &   $1.21 \pm 0.07$ &    $-19.8$  &  $-0.7$    \\
 G7    & 02:09:31.49 & $-$04:38:21.1  &   $23.68 \pm 0.04$  &  $1.1287 \pm 0.0002$ & stellar  &  $11.9$ & $ 98$  &  $ -420$ &   $1.38 \pm 0.07$ &    $-20.5$  &  $-0.4$    \\
 G8    & 02:09:30.27 & $-$04:38:16.6  &   $24.93 \pm 0.06$  &  $1.1300 \pm 0.0002$ & [O\,II]  &  $12.0$ & $ 99$  &  $ -240$ &   $1.02 \pm 0.08$ &    $-19.0$  &  $-1.0$    \\
 G9    & 02:09:30.95 & $-$04:38:14.4  &   $23.97 \pm 0.05$  &  $1.1272 \pm 0.0002$ & stellar  &  $12.1$ & $ 99$  &  $ -630$ &   $1.32 \pm 0.06$ &    $-20.2$  &  $-0.5$    \\
G10    & 02:09:31.01 & $-$04:38:13.1  &   $24.09 \pm 0.06$  &  $1.1260 \pm 0.0002$ & stellar  &  $13.5$ & $111$  &  $ -800$ &   $1.23 \pm 0.07$ &    $-20.1$  &  $-0.6$    \\
G11    & 02:09:31.48 & $-$04:38:15.8  &   $25.20 \pm 0.09$  &  $1.1299 \pm 0.0002$ & [O\,II]  &  $14.9$ & $122$  &  $ -250$ &   $1.04 \pm 0.10$ &    $-18.7$  &  $-1.1$    \\
G12    & 02:09:29.99 & $-$04:38:15.7  &   $25.53 \pm 0.11$  &  $1.1305 \pm 0.0002$ & [O\,II]  &  $15.7$ & $129$  &  $ -170$ &   $1.03 \pm 0.20$ &    $-18.4$  &  $-1.2$    \\
G13    & 02:09:31.80 & $-$04:38:21.0  &   $24.56 \pm 0.08$  &  $1.1351 \pm 0.0002$ & [O\,II]  &  $16.3$ & $134$  &  $ +510$ &   $0.98 \pm 0.09$ &    $-19.4$  &  $-0.9$    \\
G14    & 02:09:29.66 & $-$04:38:25.1  &   $24.13 \pm 0.06$  &  $1.1271 \pm 0.0002$ & stellar  &  $16.6$ & $137$  &  $ -650$ &   $1.01 \pm 0.10$ &    $-19.8$  &  $-0.7$    \\
G15    & 02:09:29.66 & $-$04:38:26.0  &   $25.04 \pm 0.10$  &  $1.1232 \pm 0.0002$ & [O\,II]  &  $16.6$ & $136$  &  $-1200$ &   $1.02 \pm 0.16$ &    $-18.9$  &  $-1.0$    \\
G16    & 02:09:31.82 & $-$04:38:34.0  &   $25.23 \pm 0.09$  &  $1.1264 \pm 0.0002$ & [O\,II]  &  $17.6$ & $145$  &  $ -750$ &   $0.90 \pm 0.21$ &    $-18.7$  &  $-1.1$    \\
G17    & 02:09:30.39 & $-$04:38:45.0  &   $25.30 \pm 0.10$  &  $1.1307 \pm 0.0002$ & [O\,II]  &  $19.6$ & $161$  &  $ -140$ &   $0.96 \pm 0.13$ &    $-18.6$  &  $-1.2$    \\
G18    & 02:09:30.71 & $-$04:38:47.6  &   $24.46 \pm 0.05$  &  $1.1299 \pm 0.0002$ & [O\,II]  &  $21.5$ & $177$  &  $ -250$ &   $1.02 \pm 0.07$ &    $-19.5$  &  $-0.8$    \\
G19    & 02:09:29.44 & $-$04:38:41.5  &   $25.61 \pm 0.07$  &  $1.1343 \pm 0.0002$ & [O\,II]  &  $25.1$ & $206$  &  $ +370$ &   $0.97 \pm 0.28$ &    $-18.3$  &  $-1.3$    \\
G20    & 02:09:31.10 & $-$04:38:00.5  &   $23.18 \pm 0.01$  &  $1.1368 \pm 0.0002$ & stellar  &  $26.1$ & $215$  &  $ +720$ &   $1.48 \pm 0.04$ &    $-21.4$  &  $+0.0$    \\
G21    & 02:09:29.15 & $-$04:38:43.5  &   $23.17 \pm 0.02$  &  $1.1259 \pm 0.0002$ & stellar  &  $29.8$ & $244$  &  $ -820$ &   $1.01 \pm 0.03$ &    $-20.8$  &  $-0.3$    \\
G22    & 02:09:31.59 & $-$04:38:54.4  &   $23.42 \pm 0.02$  &  $1.1331 \pm 0.0002$ & stellar  &  $30.8$ & $253$  &  $ +200$ &   $1.48 \pm 0.05$ &    $-21.1$  &  $-0.1$    \\
G23    & 02:09:28.85 & $-$04:38:41.1  &   $23.36 \pm 0.05$  &  $1.1259 \pm 0.0002$ & stellar  &  $32.4$ & $266$  &  $ -820$ &   $1.10 \pm 0.04$ &    $-20.7$  &  $-0.3$    \\
G24    & 02:09:28.80 & $-$04:38:40.6  &   $25.04 \pm 0.06$  &  $1.1345 \pm 0.0002$ & [O\,II]  &  $32.8$ & $269$  &  $ +390$ &   $0.93 \pm 0.08$ &    $-18.9$  &  $-1.0$    \\
G25    & 02:09:31.16 & $-$04:39:02.8  &   $23.58 \pm 0.03$  &  $1.1349 \pm 0.0003$ & stellar  &  $37.1$ & $305$  &  $ +450$ &   $1.47 \pm 0.12$ &    $-21.0$  &  $-0.2$    \\
G26    & 02:09:28.29 & $-$04:38:18.9  &   $22.36 \pm 0.01$  &  $1.1311 \pm 0.0004$ & stellar  &  $37.8$ & $310$  &  $  -80$ &   $1.40 \pm 0.10$ &    $-22.2$  &  $+0.3$    \\
G27    & 02:09:30.95 & $-$04:37:29.7  &   $23.00 \pm 0.03$  &  $1.1354 \pm 0.0004$ & [O\,II]  &  $56.5$ & $465$  &  $ +520$ &   $1.12 \pm 0.21$ &    $-21.1$  &  $-0.2$    \\
\hline
\end{tabular}
\tablenotetext{\scriptsize a}{\scriptsize \textcolor{black}{We adopt an absolute magnitude of $M_B=-21.5$ for an $L_*$ galaxy based on the luminosity function measurement from \cite{Faber:2007}. This corresponds to a luminosity of $\approx 5\times10^{10}\ L_{\rm \odot}$. }}

\end{table*}

To characterize the group environment of \FBQS, we identified 27 galaxies in our spectroscopic survey with secure redshifts and velocities within $\pm 1500$ \kms\ of the quasar systemic redshift. Table \ref{table:galaxies} summarizes the properties of these galaxies including their right ascension (R.A.), declination (decl.), apparent magnitude in the F814W filter ($m_{\rm F814W}$), redshift, redshift measurement type ([O\,II] emission or stellar absorption), projected angular ($\Delta \theta$) and physical distance ($d$) from the quasar sightline, line-of-sight velocity relative to the quasar ($\Delta v$), strength of the 4000 \AA\ break ($D_n(4000)$), rest-frame B-band absolute magnitude ($M_B$), and $B$-band luminosity relative to $L_*$ based on the $z=1.1$ luminosity function from \cite{Faber:2007}. 

To estimate the mass of the quasar host group, we measured the velocity dispersion of the group members including the quasar. We found a mean group velocity of $\Delta v _{\rm group} = -130$ \kms\ \textcolor{black}{relative to the quasar} and a velocity dispersion of $\sigma_{\rm group} \approx 550$ \kms. Assuming that the group is relaxed, this line-of-sight velocity dispersion implies a dynamical mass of $\log M_{\rm dyn}/M_\odot \approx 13.7$ using the cluster dispersion-to-mass relation from \cite{Munari:2013}. This is consistent with halo mass expectations based on the black hole mass of \FBQS\ and the black hole mass$-$halo mass relation inferred by \cite{Gaspari:2019}. The luminosity weighted group center is $\approx 60$ pkpc West and $\approx 20$ pkpc North of the quasar assuming the quasar host has a luminosity of $L_B=1-3\ L_*$ \textcolor{black}{which is typical of luminous AGN \citep[e.g.][]{Zakamska:2006}}.  \textcolor{black}{The location of the luminosity weighted group center is driven away from that of the quasar primarily by one luminous galaxy, G26, which falls outside of the MUSE field-of-view.} The MUSE galaxy redshift survey is deeper than the LDSS3 survey, so the group center may be biased toward the center of the MUSE field-of-view. A full image of the group is included in the Appendix.

\begin{figure*}
\centering
	\includegraphics[width=\textwidth, angle=0]{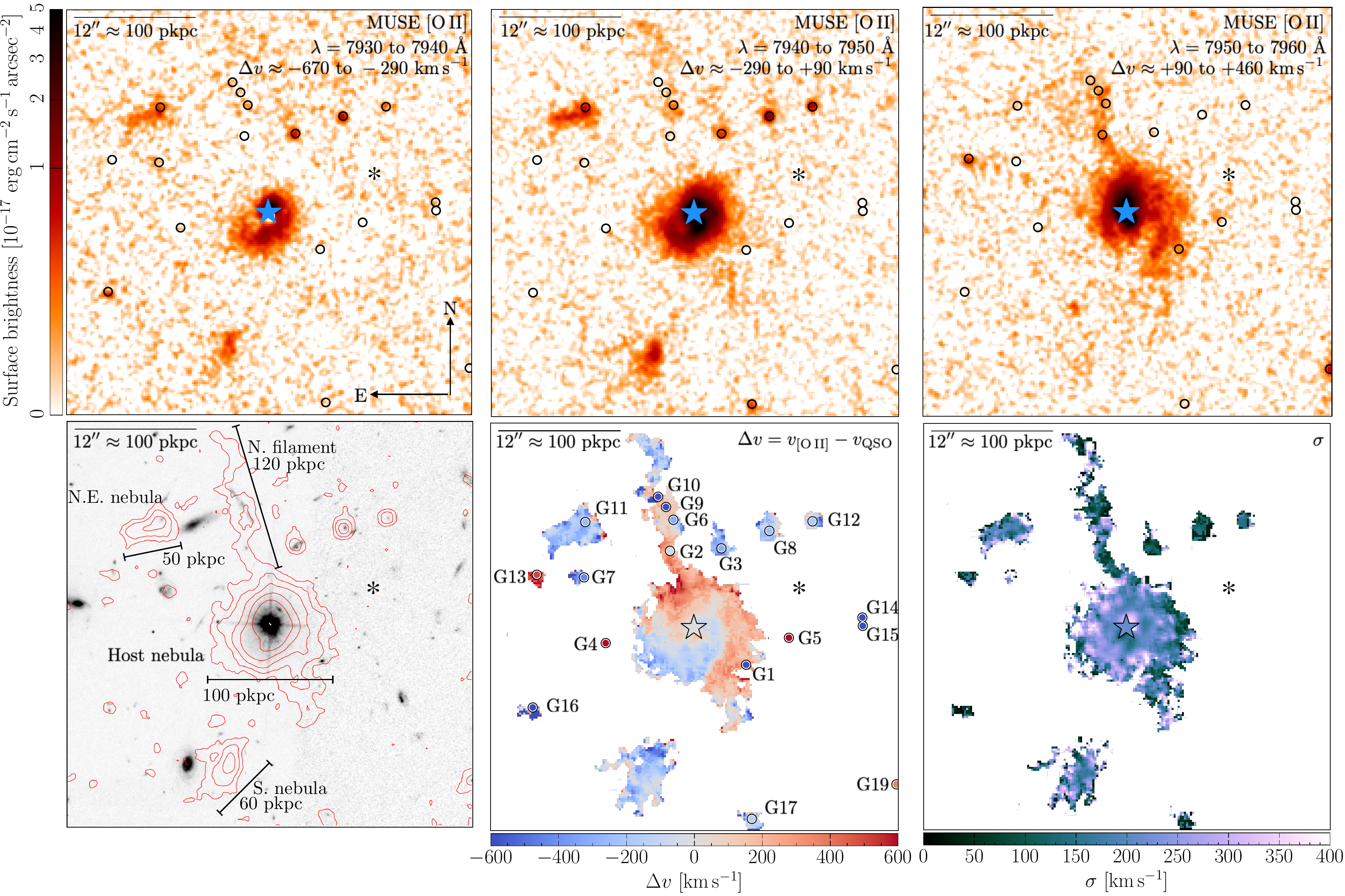}
	\caption{{\it Top panels}: Continuum-subtracted, narrow-band images around the [O\,II] doublet with wavelength ranges chosen to highlight nebulae in the environment of \FBQS. The wavelength range and corresponding approximate [O\,II] line-of-sight velocity relative to the quasar systemic are labelled in the top right corner of each panel. In each panel, the locations of galaxies in the quasar host environment are marked with black circles and the quasar position is marked by a star. {\it Bottom left panel}: [O\,II] surface brightness contours overlayed on the HST ACS/F814W image of the field marking surface brightness levels of $0.15, 0.3, 0.6, 1.2,$ and $2.4\times10^{-17}\ {\rm erg\,s^{-1}\,cm^{-2}\,arcsec^{-2}}$ after smoothing with a $1''\times1''$ boxcar. The [O\,II] contours are formed by summing the three images in the top panels. {\it Bottom middle and right panels}: [O\,II] line-of-sight velocity and line-of-sight velocity dispersion maps of the nebulae around \FBQS\ calculated with doublet fitting as described in the text. The locations of group members (quasar) are marked by black circles (black star) in the middle panel and the color interior to the symbols shows the objects systemic line-of-sight velocity relative to the quasar, $\Delta v$. The approximate luminosity-weighted group center is marked with an asterisk in all panels.
	}
	\label{figure:nebulae}
\end{figure*}

\section{Giant nebulae around \FBQS}
\label{section:nebulae}

\begin{table*}
\caption{Summary of properties of the nebulae around \FBQS}
\label{table:nebulae}
\hspace{-3cm}
\resizebox{1.17\textwidth}{!}{
\begin{tabular}{cccccrrcccc}
\hline
name & $\Delta \theta$ & $d$ & $\Delta v$ & $\sigma$ & \multicolumn{1}{c}{length-scale} & \multicolumn{1}{c}{area} & [O\,II] surface brightness\tablenotemark{a} & [O\,II] flux\tablenotemark{a} & [O\,II] luminosity\tablenotemark{a} \\
         & $\rm arcsec$ & (pkpc) & (\kms) & (\kms) & \multicolumn{1}{c}{(pkpc)} & \multicolumn{1}{c}{(pkpc$^2$)} & ($\rm erg\,s^{-1}\,cm^{-2}\,arcsec^{-2}$) & ($\rm erg\,s^{-1}\,cm^{-2}$) & $(\rm erg\,s^{-1})$\\
\hline
\hline
N.E. nebula & $15.0$ & $120$ & $-240$ & $180$ & $50$ & $900$  & $0.1$ to \ $1.0\times10^{-17}$ & $5.1\times10^{-17}$   & $3.6\times10^{41}$ \\
S. nebula & $15.0$  &  $120$ & $-250$ to $-110$ & $170$ & $60$ & $1100$ & $0.1$ to \ $1.0\times10^{-17}$ & $5.8\times10^{-17}$  & $4.1\times10^{41}$\\
N. filament & $6.0$ to $20.5$ & $50$ to $170$ & $-250$ to $+330$ & $100$ to $230$ & $120$ & $2000$ &  $0.1$ to \ $0.3\times10^{-17}$ & $5.4\times10^{-17}$ & $3.8\times10^{41}$ \\
Host nebula & $0$ to $6$ & $0$ to $50$ & $-300$ to $+540$ & $100$ to $300$ & $100$ & $6800$ & $0.1$ to $24.3\times10^{-17}$ & $1.2\times10^{-15}$\tablenotemark{a} & $8.5\times10^{42}$\\
\hline
\end{tabular}}
\tablenotetext{\scriptsize a}{\scriptsize \textcolor{black}{The total fluxes and luminosities are integrated within isophotal areas with surface brightness greater than $10^{-18}\ {\rm erg\,s^{-1}\,cm^{-2}\,{arcsec}^{-2}}$}. For the Host nebula, we masked a circular region with radius $r=1''$ when measuring the total [O II] flux and luminosity to avoid large residuals from the quasar subtraction. \textcolor{black}{If this region is not masked, the measured flux and luminosity increase by a factor of two.}}
\end{table*}

At the redshift of \FBQS, the [O\,II] $\lambda\lambda$3727, 3729 doublet, which traces cool ($T\sim10^4$ K) ionized gas, is observed at $\approx 7940$ \AA. To identify [O\,II] emitting nebulae around the quasar, we performed continuum subtraction on the quasar-light subtracted MUSE datacube by fitting low-order polynomials to each spaxel over a wavelength interval of 7800 to 8100 \AA\ after masking the 50 \AA\ region around [O\,II]. We then subtracted the continuum fit from each spaxel to produce a emission-line datacube with a typical 3-$\sigma$ detection limit of $10^{-18}\ {\rm erg\,s^{-1}\,cm^{-2}\,arcsec^{-2}}$ for a line with a width $\sigma=200$ \kms\ averaged over an $r=0.5''$ aperture.  

The continuum subtracted datacube reveals the presence of four distinct, $50-100$ pkpc scale ionized nebulae emitting in [O\,II] in the quasar host group and several smaller-scale nebulae closely associated with group members. To visualize these nebulae, the top three panels of Figure \ref{figure:nebulae} display continuum subtracted [O\,II] surface brightness maps integrated over three 10 \AA\ intervals which correspond to line-of-sight velocities of $\Delta v \approx -670$ to $-290$, $-290$ to $90$, and $90$ to $460$ \kms\ relative to the quasar systemic velocity. We note that the [O\,II] doublet separation corresponds to $\approx 200$ \kms\, so this velocity correspondence is approximate. Throughout the paper, we refer to these nebulae by their morphology and location relative to the quasar as the Northeast (N.E.) nebula, the South (S.) nebula, the North (N.) filament, and the Host nebula (see Figure \ref{figure:nebulae}). \textcolor{black}{In addition to these, there is a possible second filamentary feature extending south of the Host nebula seen in Figure \ref{figure:nebulae}. However it is less prominent than the N. filament and may be an extension of the Host nebula's arm-like feature}.

To visualize the morphologies of the nebulae relative to galaxies in the group, the bottom left panel of Figure \ref{figure:nebulae} displays the {\it HST} image of the field overlaid with [O\,II] surface brightness contours computed from the sum of the three images shown in the top three panels. The nebulae are labeled by their name and approximate length scale in the bottom-left panel. To better quantify the kinematics of these nebulae, we also performed [O\,II] doublet fitting to the datacube as described in \cite{Johnson:2018} and \cite{Helton:2021}. The bottom middle and bottom-right panels of Figure \ref{figure:nebulae} display the line-of-sight velocity relative to the quasar and line-of-sight velocity dispersion (corrected for the MUSE line spread function \citep[][]{Bacon:2017}), respectively. In most cases, the emission is too broad or too low in S/N to measure the [O\,II] 3729-to-3727 doublet ratio, which can vary between 0.35 and 1.5. Uncertainty in the doublet ratio introduces a systematic uncertainty of $\approx 50-80$ \kms\ in the [O\,II] velocity centroid. Table \ref{table:nebulae} summarizes the properties of the nebulae,  and we discuss each in turn in the following.

\subsection{Northeast nebula}
The N.E. nebula is located $\Delta \theta \approx 15''$ or $d\approx 120$ pkpc from \FBQS, has a length scale of $\approx 50$ pkpc and [O\,II] surface brightness ranging from $\approx0.1$ to $1.0\times10^{-17}\ {\rm erg\,s^{-1}\,cm^{-2}\,arcsec^{-2}}$. The N.E. nebula exhibits a head-tail morphology oriented approximately East-West with a surface brightness peak that is spatially coincident with a group member galaxy, G11. Together, the morphology and spatial coincidence with G11 suggest that the N.E. nebula arises from ram pressure stripping of the ISM of G11 as it moves through the hot halo of the quasar host group. Such ``jelly-fish'' galaxies are often observed in \textcolor{black}{galaxy} clusters and groups \citep[e.g.][]{Fumagalli:2014, Poggianti:2017, Chen:2019a, Boselli:2019} and around quasar hosts at $z\approx 0.5$ \citep[e.g.][]{Johnson:2018, Helton:2021} \textcolor{black}{where tidal stripping also plays a role \citep[see also ][for a case of likely tidal stripping around a quasar at $z=6$]{Decarli:2019}}. We caution that the continuum S/N of G11 is too low to measure a stellar-absorption-based redshift.

\subsection{South nebula}
The S. nebula is located $\Delta \theta = 15''$ or $d\approx 120$ pkpc South of \FBQS, extends over a length-scale $\approx 60$ pkpc, and exhibits [O\,II] surface brightness of $\approx0.1$ to $1.0\times10^{-17}\ {\rm erg\,s^{-1}\,cm^{-2}\,arcsec^{-2}}$.
While the surface brightness contours of the S. nebula are somewhat elongated, its surface brightness peak is not coincident with any continuum sources in the {\it HST} image. At the location of S. nebula, the {\it HST} image is sensitive to galaxies of $M_{B}=-16$ at $z=1.13$. The morphology and lack of associated galaxies suggest that the S. nebula is a collection of cool intragroup medium clouds in the massive halo \citep[e.g.][]{Nelson:2020}, similar to others observed around quasars \citep[e.g.][]{Johnson:2018, Helton:2021} at $z\approx 0.5$. However, we note that the S. nebula could also represent ram pressure debris stripped from the ISM of a dwarf galaxy fainter than $L_B\approx 0.006 L_*$. \textcolor{black}{The brightest galaxy without a robust redshift near the S. nebula falls along the Northern edge of faintest [O\,II] surface brightness contour shown in the bottom left panel of Figure \ref{figure:nebulae}. This galaxy has an apparent magnitude of $m_{\rm F814W} = 26.0$ which would correspond to an absolute magnitude of $M_B = -18.1$ at $z=1.13$, but it lacks the strong nebular emission expected from a galaxy experiencing on-going ram pressure stripping.}

\subsection{North filament}

The N. filament extends from $\Delta\theta \approx 20.5''$ or $d=170$  pkpc North of the quasar toward it and intersects with the Host nebula $6''$ or $d=50$ pkpc from the quasar. Despite its length, the N. filament is narrow with a width that ranges from $1''$ to $3''$ or $\approx 8$ to $24$ pkpc. The N. filament is fainter than the other nebulae in the field with a peak [O\,II] surface brightness of $0.3\times10^{-17}\ {\rm erg\,s^{-1}\,cm^{-2}\,arcsec^{-2}}$. The kinematics of the N. filament are complex, and vary from $\Delta v \approx -250$ \kms\ at its northern-most points to $\Delta v \approx +350$ \kms\ where it intersects with the Host nebula.
The N. filament is spatially coincident with nearby galaxies, extending $\approx 30$ pkpc to the North of G10 and connecting G9, G6, and G2 before intersecting with the Host nebula. Kinematically, the velocity of the N. filament appears distinct from G2, G6, G9, and G10 but matches that of the Host nebula where they intersect, as seen in Figure \ref{figure:nebulae}.

\textcolor{black}{Optically emitting outflows from radio-loud AGN are commonly observed with orientations aligned with radio lobes \citep[e.g.][]{Nesvadba:2017a} which could explain the morphology of the N. filament. However, these outflows typically exhibit broad line widths of $500<{\rm FWHM}<1500$ \kms\ while the N. filament exhibits a median width of FWHM $\approx 300$ \kms. Furthermore, radio observations from FIRST \citep[][]{Becker:2001}, VLASS \citep[][]{Lacy:2020}, and the XXL Survey GMRT 610 Mhz continuum observations \citep[][]{Smolcic:2018} reveal no evidence of a jet or lobe aligned with the N. filament despite $3\sigma$ detection limits of  450, 210, and 140 $\mu$Jy per beam respectively. Starburst driven outflows and radio-quiet AGN driven outflows typically exhibit wide opening angles \citep[e.g.][]{Liu:2013a, Rupke:2019, Burchett:2021, Zabl:2021} inconsistent with the morphology of the N. filament. Together, the kinematics, morphology, and lack of detected jets disfavor an outflow origins for the N. filament.}

\begin{figure}
\includegraphics[width=0.45\textwidth]{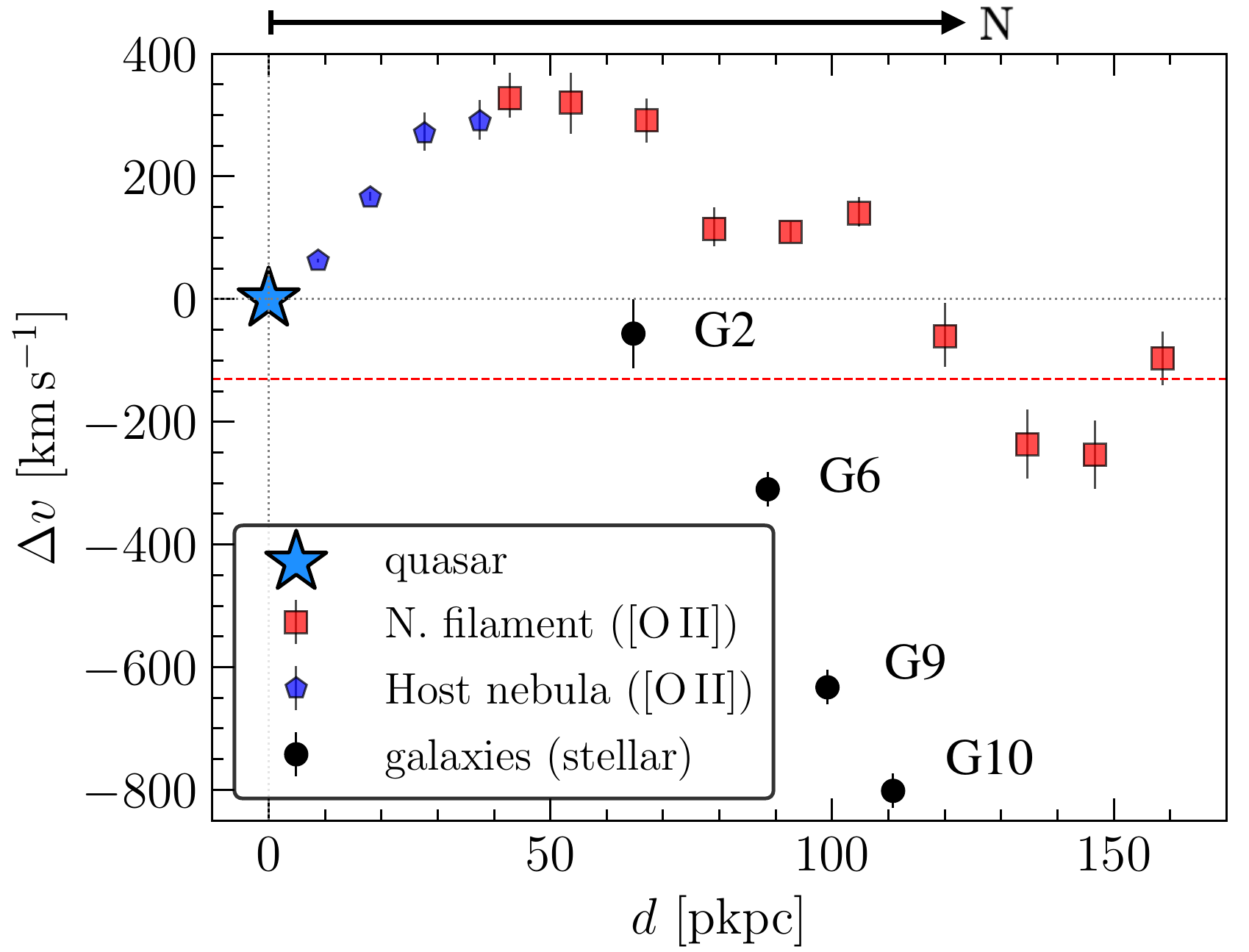}
\caption{Line-of-sight velocity of the Host nebula and N. filament as a function of projected distance from the quasar compared with nearby galaxies. Points corresponding to the Host nebula are shown as blue pentagons and those corresponding to the N. filament are shown as red squares. The N. filament measurements are made following the filament contours seen in Figure \ref{figure:nebulae} with $0.6-1.2''$ radius extraction apertures depending on the width of the filament at each location. The measurements for the Host nebula are made moving North from the quasar toward the N. filament with $0.6"$ radius extractions. Galaxies G2, G6, G9, and G11 overlap spatially with the N. filament and their stellar velocities vs projected distances are shown in black circles in the bottom panel for comparison with the nebulae the quasar is marked by a blue star. Despite their spatial coincidence seen in Figure \ref{figure:nebulae}, the line-of-sight [O\,II] velocities of the N. filament differ from nearby galaxies by $\gtrsim 300$ \kms. \textcolor{black}{The luminosity-weighted group velocity is indicated by the red dashed line.}}
\label{figure:kinematics}
\end{figure}

The morphology of the N. filament can be explained if it arises from \textcolor{black}{overlapping nebulae resulting from on-going} ram-pressure stripping of group members or cool, filamentary accretion. The \textcolor{black}{on-going} ram pressure stripping scenario can explain the spatial coincidence with galaxies but would require a somewhat contrived chance alignment. Further, the significant differences between the \textcolor{black}{[O\,II] emission velocities of the nebulae vs. stellar absorption velocities of galaxies} G2, G6, G9, and G10 disfavors ongoing stripping of ISM. To quantify the velocity differences, Figure \ref{figure:kinematics} displays the [O\,II]-based velocities for the N. filament and the stellar absorption-based velocities of G2, G6, G9, and G10 versus projected distance from the quasar. The galaxy and nebular line-of-sight velocities differ by more than $300$ \kms, \textcolor{black}{even at the locations of the galaxies}. \textcolor{black}{This mis-match is inconsistent with previous observations of nebulae arising from ongoing ram pressure stripping of ISM \citep[][]{Johnson:2018, Boselli:2019, Chen:2019a, Helton:2021}.}

The long, narrow morphology, lack of kinematic correspondence with galaxies, and matching velocity to the northern edge of the Host nebula all suggest the N. filament arises from cool, filamentary accretion. Such cool accretion could result from several mechanisms \textcolor{black}{including (1) a cooling flow from the hot halo of the quasar host group, (2) cool, filamentary gas accreting from the IGM, or (3) accreting cool CGM and intragroup medium from a less massive galaxy group as it falls into the quasar host system. Filamentary cooling flows in cool core clusters are observed in optical emission lines including [O\,II] \citep[e.g.][]{McDonald:2010, McDonald:2012a}. In this case, the N. filament would be the longest and highest redshift known such flow despite arising in a less massive system with a velocity dispersion $\approx 3\times$ lower than that of the current record holder, the Phoenix Cluster \citep[][]{McDonald:2012a}. Spatial coincidence with group members is not typical for cooling flows in  cool-core clusters. Instead, the filament could represent chaotic cold accretion from the hot halo \citep[e.g.][]{Gaspari:2018} if interactions between G2, G6, G9, G10, and the quasar host induce turbulenct cooling. In this case, the complex velocity shear would reflect bulk motion of the hot halo.}

\textcolor{black}{
Alternatively, the N. filament's morphology and coincidence with galaxies can be explained by cool accretion of an intergalactic medium filament connecting galaxies G2, G6, G9, and G10 or CGM from a galaxy group containing G2, G6, G9, and G10 as the group is accreted by the quasar host system.  In both cases, the velocity difference between the nebulae and nearby galaxies can be explained if the gas experienced ram pressure deceleration over sufficiently long timescales. Bulk motion of hot halos can include rotation \citep[e.g.][]{Hodges-Kluck:2016, Oppenheimer:2018} or other complex patterns which could explain the velocity shear of the N. filament.} \textcolor{black}{The accreting CGM scenario would require tidal forces to explain the elongated morphology.}

An IGM filament would explain the morphology and potentially connect galaxies. \textcolor{black}{The ratio of the stream minor axis radius to the estimated host halo virial radius ($\approx 560$ pkpc) is $\approx 1-4\%$, consistent with predictions for cool inflowing IGM streams from \cite{Mandelker:2020}}.  The N. filaments properties can also be explained by CGM or intragroup medium in the presence of interactions. In particular, if the filament originated as CGM or intragroup medium around G2, G6, G9, and G10, the elongated morphology could be the result of tidal and ram pressure forces experienced as the group members and surrounding gas fell toward the quasar host.
In summary, the morphology and kinematics of the N. filament are \textcolor{black}{consistent with} cool filamentary accretion from the IGM, from the CGM/intragroup medium around a group of galaxies being accreted by the quasar host system, or interaction-induced cooling of hot gas.

\subsection{Host nebula}

\begin{figure*}
\includegraphics[width=\textwidth]{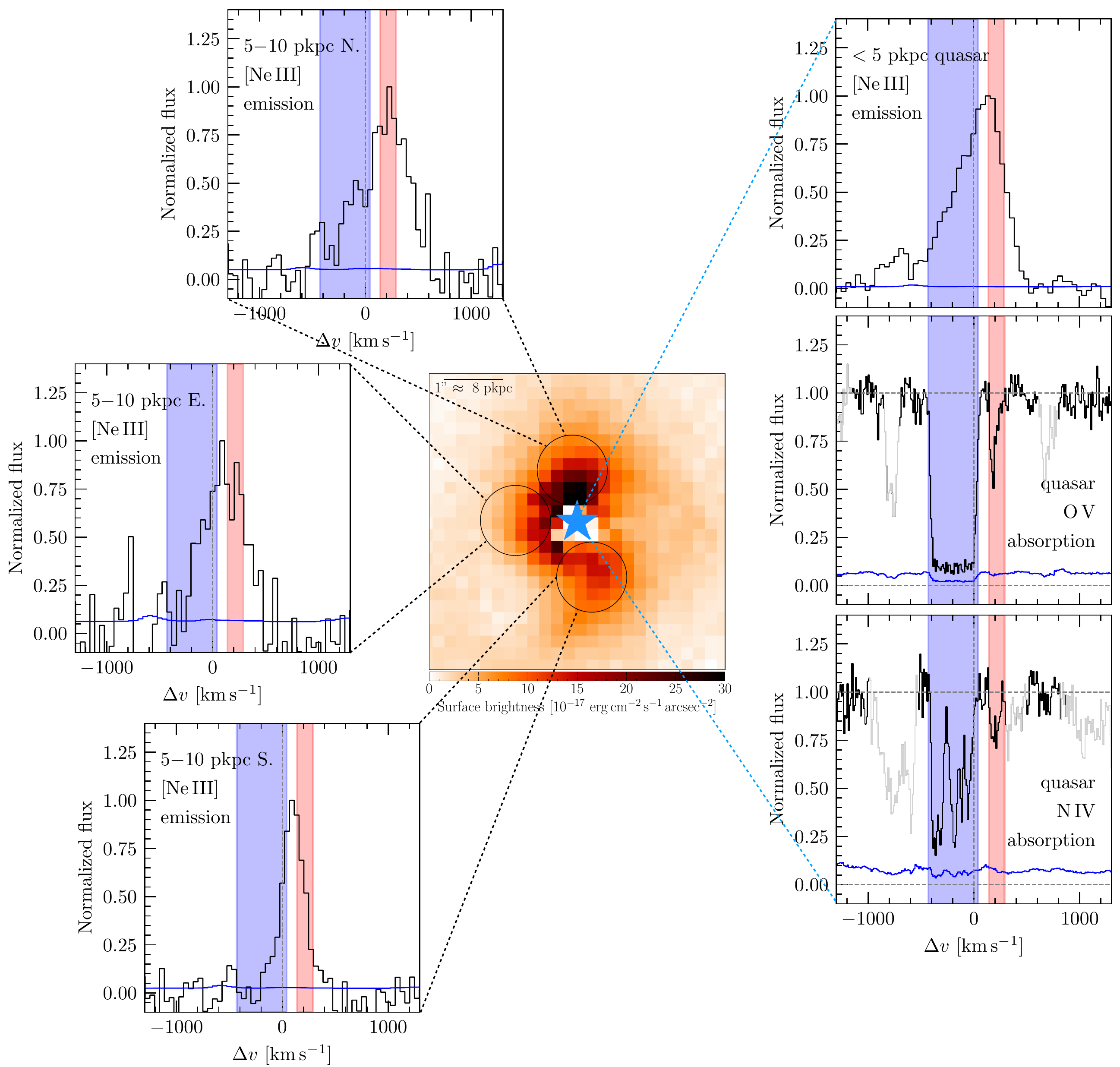}
\caption{\textcolor{black}{Summary of emitting and absorbing gas kinematics near the nucleus. The image in the {\it center panel} displays the [O\,II] surface brightness map zoomed-in and scaled to highlight structures near the nucleus. The quasar position is marked by a blue star and pixels touching the star are subject to large uncertainties in quasar light subtraction. The {\it top right} panel shows normalized flux of [Ne\,III]  versus line-of-sight velocity from the quasar while the three panels on the {\it left} display [Ne\,III] emission for circular apertures offset from the quasar centroid. The regions corresponding to each spectrum are shown in black circles and connected to the corresponding spectral panel by a dashed black line.  The {\it bottom} two panels on the right display O\,V (middle right) and N\,IV (bottom right) narrow, associated absorption from the COS FUV spectrum of \FBQS. The COS spectrum reveals both outflowing (blue-shifted) and inflowing (redshifted) absorption and photoionization analysis of the absorbers from \cite{Finn:2014} indicates that the absorbing gas arises at distances of ~2$-$6 pkpc from the nucleus. For reference, the velocity ranges of the outflowing and inflowing material observed in the quasar absorption spectrum are shown in blue and red highlight respectively. The inflowing component exhibits a line-of-sight velocity of $\approx+200$ \kms\ that is consistent with the velocity field seen in [Ne\,III] emission seen at $5-10$ kpc from the quasar in the panels on the left. The outflowing absorbing gas shows velocities that are comparable to the nuclear [Ne\,III] but exceed the blueshifts seen in emission at $d=5-10$ kpc seen in the left three panels.}}
\label{figure:inner}
\end{figure*}

The Host nebula is approximately centered on the quasar and extends to a radius of $\Delta \theta=6''$ or $d\approx 50$ pkpc from the quasar with [O\,II] surface brightness levels ranging from $0.1$ to $24\times10^{-17}\ {\rm erg\,s^{-1}\,cm^{-2}\,arcsec^{-2}}$. The kinematics of the Host nebula are complex with line-of-sight velocities ranging from $\approx-300$ to $+540$ \kms\ and distinct spiral-like structure seen in Figure \ref{figure:nebulae}.

The radial extent, morphological correspondence with the quasar, and high peak surface brightness, suggest that the Host nebula represents extended, ionized ISM and \textcolor{black}{diffuse gas} around the quasar host. However, the Host nebula's kinematics are more complex than the canonical ``spider'' diagram expected for rotating disks. \textcolor{black}{Quantitative analysis of the velocity structure function of the Host nebula demonstrates that it follows expectations for Kolmogorov turbulence with isotropic, homogeneous, and incompressible gas \citep[see][]{Chen:2022}}.

The unusual spiral structure visible in the surface brightness panels (top) and velocity map (bottom middle) in Figure \ref{figure:nebulae} may be a signature of tidal arms from a past interaction, possibly one that helped fuel the quasar. However, the bright quasar limits our ability to search for interaction signatures that would confirm this scenario.

To investigate whether there is a significant, large-scale inflow or outflow associated with the Host nebula, we take advantage of the archival COS spectrum of the quasar, which enables clean differentiation of outflowing (blueshifted) and inflowing (redshifted) gas along the quasar sightline. The COS spectrum reveals narrow associated absorption, including both outflowing and inflowing components detected in an array of ions. The middle and bottom panels on the right of Figure \ref{figure:inner}  show O\,V $\lambda 629$ and N\,IV $\lambda 569$ absorption as a function of line-of-sight velocity relative to the quasar. \cite{Finn:2014} conducted ionization analysis of these absorbers and found that the absorbing clouds likely arise at distances of $\gtrsim 2$ pkpc from the nucleus. This is only slightly smaller than the angular resolution of the seeing limited MUSE data ($0.7''$ corresponds to $\approx 6$ pkpc at $z=1.13$), providing a unique opportunity to jointly study the emitting and absorbing gas near the quasar.

\textcolor{black}{The regions of the Host nebula near the quasar are bright enough to be observed in [Ne\,III] $\lambda 3869$ emission, which enables more precise velocity measurements. To search for kinematic correspondence between the emitting gas and associated absorbers, Figure \ref{figure:inner} displays peak-normalized [Ne\,III] emission as a function of line-of-sight velocity for a central ($d<5$ pkpc) extraction centered on the quasar in the top right panel and extractions at $d=5-10$ pkpc immediately to the N. (left/top panel), E. (left/middle panel), and S. (left/bottom panel).}

The [Ne\,III] emission at $d<5$ pkpc from the quasar sightline peaks between the quasar systemic velocity and that of the inflowing component seen in associated absorption in O\,V, N\,IV, and other ions. The circum-nuclear [Ne\,III] emission also exhibits a prominent blue wing extending from $\Delta v \approx -200$ \textcolor{black}{to $-400$ \kms}, similar to the outflowing absorbers. The similarity in \textcolor{black}{velocity range} observed between the associated inflow/outflow observed in absorption and the circumnuclear [Ne\,III] suggests that the emissions may trace \textcolor{black}{different phases and locations} of the the same gas flows traced in absorption.

Further from the nucleus, the [Ne\,III] emission at $d=5-10$ pkpc shows decreased prominence of the blue wing, suggesting that the \textcolor{black}{faster outflow component at $\Delta v < -200$ \kms } may be confined to central regions of the host.
On the other hand, the redshifted emission component is more prominent and peaks near the associated inflow velocity, particularly \textcolor{black}{in the N. extraction}. The more extended nature of the redshifted emission and kinematic coincidence with the associated inflow suggest a common origin and a relation to Host nebulae. \textcolor{black}{In this case, the emitting gas near the nucleus would have to be in front of the quasar and oriented towards us within the quasar ionizing radiation cone to be illuminated and for the line-of-sight velocity to approximately match the radial velocity of the down-the-barrel inflow}.

\section{Discussion and Conclusions}
\label{section:conclusions}

Based on a combination of deep MUSE observations and archival HST data, we discovered four distinct and giant ionized nebulae in the environment of \FBQS, a luminous quasar at $z=1.13$. Some of these nebulae are likely related to an inflow detected in UV absorption. Two of the nebulae are well separated from the quasar host galaxy and likely arise from on-going ram pressure stripping of a group member (N.E. nebula) and cool clouds in the intragroup medium (S. nebula), extending observations of large-scale streams observed in groups and around quasar hosts to higher redshift \citep[e.g.][]{Hess:2017, Johnson:2018, Helton:2021}. 

The two other giant nebulae include a $\approx 120$ pkpc filament extending to the N. of the quasar (N. filament), which intersects--both spatially and kinematically--with a $100$ pkpc diameter nebula (Host nebula) surrounding the quasar host itself. Immediately around the quasar, the Host nebula exhibits velocities similar to inflowing absorbing gas observed in UV spectra of the quasar. The morphology and kinematics of the N. filament and Host nebula and coincidental inflowing gas constitute strong evidence of large-scale, cool filamentary accretion from halo scales into the quasar host and toward the nucleus. 

If the [O\,II] emitting gas is in pressure equilibrium with a hot halo, we can gain insights into its physical conditions. \textcolor{black}{Hydrodynamical simulations predict global pressure equilibrium, although small-scale fluctuations from subsonic turbulence can occur \citep[e.g.][]{van-de-Voort:2012, Gaspari:2014}, \textcolor{black}{though observational results are mixed \citep[see][]{Werk:2014, Stern:2016, Zahedy:2019, Butsky:2020, Qu:2022}.}} Adopting the group mass from Section \ref{section:quasar} and the \textcolor{black}{generalized NFW} pressure profile from \cite{Arnaud:2010}, we estimate hot halo pressure of $P_{\rm hot}/k_{\rm B}\approx 5\times10^4\ {\rm K\, cm^{-3}}$ at \textcolor{black}{$\approx 100$ pkpc} for the N. filament. 
\textcolor{black}{Based on Cloudy v17.03 \citep[][]{Ferland:2017} photoionization equilibrium models for gas illuminated by the quasar \textcolor{black}{which dominates over the expected UV background at these distances \citep[e.g.][]{Faucher-Giguere:2019}}, we expect a temperature of $\log T/{\rm K}=4-4.5$ and a density of $n_e\approx 1-5$  ${\rm cm^{-3}}$}.

Estimating the surface brightness of [O\,II] given a density and total ionized column, $N({\rm H\,II})$, is complicated by the unknown ionization state and metallicity of the gas. However, we expect a line ratio of [O\,II]/H$\alpha\sim 1$ over a fairly wide range of conditions \textcolor{black}{for AGN photoionized gas} \citep[e.g.][]{Groves:2004}, enabling us to roughly estimate the ionized gas column because H$\alpha$ surface brightness can be approximated as ${\rm SB_{H\alpha}}\approx 1.7\times10^{-16}C \left(\frac{\langle n_e\rangle }{{\rm cm^{-3}}}\right) \left( \frac{N({\rm H\,II})}{3\times10^{20} {\rm cm^{-2}}}\right) \left( \frac{1}{1 + z}\right)^4$ $\rm erg\,s^{-1}\,cm^{-2}\,arcsec^{-2}$ where $C$ is the clumping factor, $C=\langle n_e^2\rangle/\langle n_e \rangle^2$. Assuming $n_e\sim 3\ {\rm cm^{-3}}$ and $C\approx 1$, the surface brightnesses of the N. filament corresponds to approximate ionized gas columns of $N({\rm H\,II})\sim 10^{19}\ {\rm cm^{-2}}$. \textcolor{black}{With the same assumptions, we estimate total ionized gas masses of $M_{\rm g}\sim 10^9$ and $2\times10^{10}\ M_\odot$ for the N. filament and Host nebula, respectively based on the total line luminosity following \cite{Greene:2011}. These are significantly lower than the estimates of ionized gas mass assuming the same density and near unity volume filling factor which results in $M_{\rm gas}\sim 10^{12} M_\odot$  for the N. filament assuming a cylinder with length of $l=100$ pkpc and radius of $r=10$ pkpc and $M_{\rm gas} \sim 10^{13} M_\odot$ for the Host nebula assuming a uniform sphere with radius $r=50$ pkpc. These volume-based estimates significantly exceed not only the previous luminosity-based estimates but also the expected baryon budget for the group, requiring lower mean density and higher clumping factor \citep[e.g.,][]{Cantalupo:2019}. With a clumping factor of $C\approx 100$, the discrepancy between the two mass estimates can be resolved. Insights into the density structure of the gas require new emission observations.}

\textcolor{black}{Given its low surface brightness and possible column density}, the N. filament may arise in gas analogous to Lyman limit systems and damped \lya\ absorbers but in a more overdense, higher pressure environment and subjected to intense ionizing radiation from the quasar. The $\approx80\times$  higher peak surface brightness of the inner regions of the Host nebula \textcolor{black}{suggests} significantly higher density or ionized gas column, consistent with ISM.
The inflowing absorption detected in the spectrum of \FBQS\ combined with the morphology of the extended, emitting gas of the Host nebula provides an opportunity to estimate the accretion rate in a quasar host \textcolor{black}{assuming that the absorbing gas and extended emission are tracing different phases and locations along a coherent gas flow}. Following \cite{Weiner:2009}, we assume a thin, spherical shell which results in an inflow rate of $\frac{dM}{dt} \sim \left(22\  M_{\odot}\,{\rm yr^{-1}}\right) \left( \frac{N({\rm H})}{10^{20}\,{\rm cm^{-2}}}\right) \left( \frac{R}{5\, {\rm pkpc}} \right) \left(\frac{v}{300\, {\rm km\,s^{-1}}} \right)$ where $N({\rm H})$ is the total hydrogen column density of the inflow, $R$ is the radius of the shell, and $v$ is the inflow velocity. \textcolor{black}{While it is an oversimplification,} we chose the thin-shell for ease of comparison with previous results \citep[e.g.][]{Arav:2013} \textcolor{black}{and to place a conservative upper limit on the inflow rate}. Adopting a uniform, spherical flow of radius $R$ would reduce the inflow rate by a factor of three. With a total column of $N({\rm H})\sim 10^{17}\, {\rm cm^{-1}}$ based on the ionization analysis of \cite{Finn:2014}, a radius of $R<50$ pkpc based on the maximum observed extent of the Host nebula, and the observed inflow velocity of $v\approx 150$ \kms, we infer an upper limit on the inflow rate of $\frac{dM}{dt}<0.1\ M_\odot\,{\rm yr}^{-1}$. \textcolor{black}{The inflowing absorbing gas column inferred by \cite{Finn:2014} is several orders-of-magnitude below our initial surface-brightness based estimate for the ionized column. This can be resolved if the emitting gas traces denser, higher column density gas phase with lower covering factor than the absorbing gas or equivalently, by a clumping factor of $\approx 100$ as previously suggested.} \textcolor{black}{The H\,I column for the inflow inferred by \cite{Finn:2014} is based on metal absorption only and assumes approximately solar metallicity due to unavailability of an H\,I column measurement.}

\textcolor{black}{If the absorbing gas arises closer to the nucleus at $\approx 2$ pkpc \cite[see][]{Finn:2014}, then the inflow rate estimate decreases to  $\frac{dM}{dt}<0.004\ M_\odot\,{\rm yr}^{-1}$}. \textcolor{black}{An inflow rate of  $0.1\ M_\odot\,{\rm yr}^{-1}$} corresponds to a luminosity of $\sim 5\times10^{45}\, {\rm erg\,s^{-1}}$ or a radiative luminosity of $\sim 5\times10^{44}\, {\rm erg\,s^{-1}}$ assuming a radiative efficiency of 10\%. This is several order-of-magnitude below the observed radiative luminosity of the quasar indicating a significant difference in accretion rate \textcolor{black}{at $\gtrsim 2$ pkpc compared to nearer the accretion disk. This large difference suggests highly anisotropic or highly time variable accretion. Time variable accretion will be reflected in luminosity variability of the quasar, though we note that \FBQS\ exhibits 5\% level variability in the UV on month timescales \citep[e.g.][]{Punsly:2016}. Accretion mechanisms such as chaotic cold accretion \citep[e.g.][]{Gaspari:2018} and interaction induced quasar activity \citep[e.g.][]{Goulding:2018} are expected to produce significant accretion rate variability on timescales comparable to the dynamical time at a few pkpc.} \textcolor{black}{Alternatively, the observations of accretion rate at $\gtrsim 2$ pkpc than is orders-of-magnitude lower than that inferred for the quasar engine itself on much smaller scales could be a signature of effective AGN feedback heating the CGM to slow accretion.}

\textcolor{black}{The discovery and morphokinematic analysis of multiple large nebulae including a $\approx 100$ pkpc long filament and connected $\approx 100$ kpc diameter nebulae around a luminous quasar with an inflowing associated absorber demonstrate the unique insights enabled by observations of non-resonant emission lines with wide-field IFS when coupled with down-the-barrel absorption spectroscopy. Developing a better understanding of the origins, fate, and physical conditions of these gas flows requires observations of emission lines from a wider variety of ions which will become possible with upcoming near-IR IFS such as MIRMOS on Magellan \citep[][]{Konidaris:2020} and HARMONI \citep[][]{Thatte:2021} on the E-ELT.}

\section*{Acknowledgements}
The authors are grateful for the efforts of the anonymous referee whose thorough feedback strengthened the paper. SDJ gratefully acknowledges partial support from HST-GO-15280.009-A, HST-GO-15298.007-A, HST-GO-15655.018-A, and HST-GO-15935.021-A. MG acknowledges partial support by HST GO-15890.020/023-A, the {\it BlackHoleWeather} program, and NASA HEC Pleiades (SMD-2359).. SC gratefully acknowledges support from the European Research Council (ERC) under the European Union’s Horizon 2020 research and innovation programme grant agreement No 864361. GLW acknowledges partial support from HST-GO-15655.006-A and HST-GO-16016.002

Based on observations from the European Organization for Astronomical Research in the Southern Hemisphere under ESO (PI: Schaye, PID: 094.A-0131) and the NASA/ESA Hubble Space Telescope (PI: Abraham, PID: 9760; PI: Morris, 12264; PI: van der Marel, PID: 12564; PI: L. Straka; PID: 14660). The COS spectra and ACS images analyzed in this paper can be accessed via \dataset[COS DOI]{https://doi.org/10.17909/4dnz-y880} and \dataset[ACS DOI]{https://doi.org/10.17909/tekk-dx10}, respectively.

The This paper includes data gathered with the 6.5 meter Magellan Telescopes located at Las Campanas Observatory, Chile.
The paper made use of the NASA/IPAC Extragalactic Database, the NASA Astrophysics Data System, Astropy \citep[][]{Astropy-Collaboration:2018}, and Aplpy \citep[][]{Robitaille:2012}. 

\appendix{}
\section{Supplemental information on the group hosting \FBQS}
\begin{figure*}
\centering
	\includegraphics[width=\textwidth, angle=0]{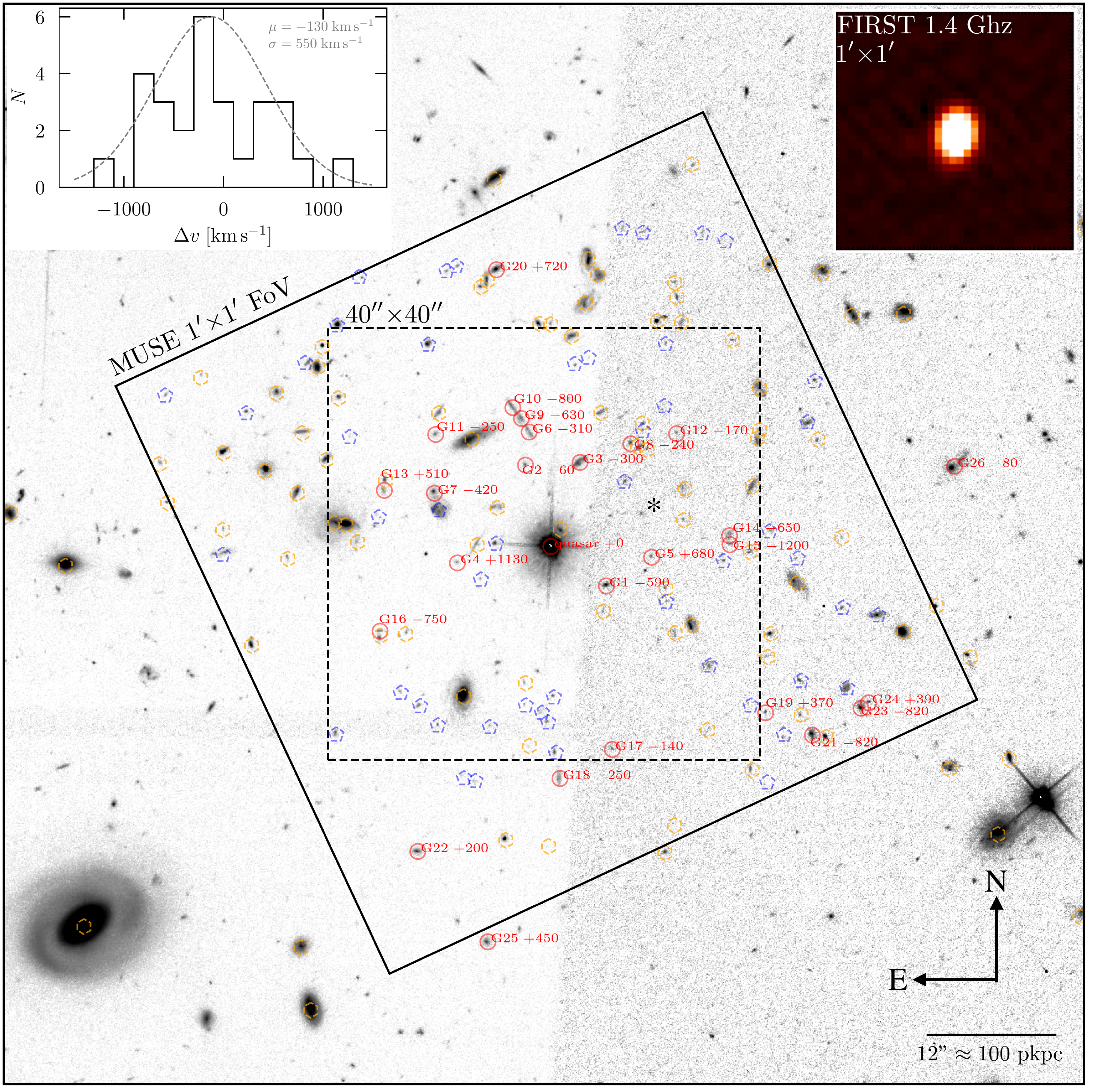}
	\caption{Summary of the redshift survey and galaxy group hosting \FBQS. The greyscale image displays the reduced, coadded {\it HST} ACS$+$F814W image of the field centered on \FBQS\ with orientation and scale marked in the bottom right. The full $1'{\times}1'$ MUSE FoV is marked by the black square with solid outline, and the $40''{\times}40''$ region shown in Figure \ref{figure:nebulae} is marked in dashed line. Galaxies in the group hosting \FBQS\ are labelled with red circles and labelled by their ID and velocity in \kms\ from Table \ref{table:galaxies}. The approximate luminosity-weighted group center is marked with an asterisk. Galaxies with secure redshifts but foreground and background to the quasar host group are marked in blue and orange dashed symbols, respectively. The inset panel on the top right displays the line-of-sight velocity histogram for group members, including \FBQS, in black, solid line while the best-fitting Guassian with mean line-of-sight velocity of $\mu=-130$ \kms\ and line-of-sight velocity dispersion of $\sigma = 550$ \kms\ is shown in dashed grey line. The 1.4 Ghz radio image from the FIRST survey is shown in the top right inset panel and confirms the core-dominated nature of this radio-loud quasar.}
	\label{figure:appendix}
\end{figure*}

To provide additional context on the galaxy group hosting \FBQS, Figure \ref{figure:appendix} displays a wider FoV version of the {\it HST} ACS$+$F814W image with galaxies labelled by group membership based on spectroscopic redshifts as described in Section \ref{section:quasar}. The figure also highlights the velocity dispersion of the group and best-fit Gaussian and the core-dominated nature of the radio component of \FBQS\ in inset panel on the top left and top right, respectively.

\bibliographystyle{aasjournal}
\bibliography{/Users/seanjoh/Dropbox/library/astro.bib} 

\end{document}